\documentclass[final,3p,times,twocolumn]{elsarticle}

\usepackage{graphicx}

\usepackage{amssymb}
\usepackage{amsmath}

\journal{Astroparticle Physics}

\begin{document}

\begin{frontmatter}



\title{Searches for  solar-influenced radioactive decay anomalies  \\ using Spacecraft RTGs}

\author[Wabash,Purdue]{D. E. Krause\corref{cor1}}
\ead{kraused@wabash.edu}
\author[Purdue Aero]{B. A. Rogers}
\author[Purdue]{E. Fischbach}
\author[Wittenberg]{J.~B.~Buncher}
\author[JPL]{A. Ging}
\author[Purdue Nuclear,Purdue]{J.~H.~Jenkins}
\author[Purdue Aero]{J.~M. Longuski}
\author[JPL,Purdue Aero]{N.~Strange}
\author[Stanford]{P. A. Sturrock}

\address[Wabash]{Department of Physics, Wabash College, Crawfordsville, IN 47933, USA}
\address[Purdue]{Department of Physics, Purdue University, 525 Northwestern Avenue, West Lafayette, IN 47907, USA}
\address[Purdue Aero]{School of Aeronautics and Astronautics, Purdue University, 701 W. Stadium Ave., West Lafayette, Indiana 47907, USA}
\address[Wittenberg]{Department of Physics, Wittenberg University, Springfield, Ohio 45501, USA}
\address[JPL]{Jet Propulsion Laboratory, California Institute of Technology, 4800 Oak Grove Drive,  Pasadena, CA  91109, USA}
\address[Purdue Nuclear]{School of Nuclear Engineering, Purdue University, 400 Central Dr., West Lafayette, IN 47907, USA}
\address[Stanford]{Center for Space Science and Astrophysics, Varian 344, Stanford University, Stanford, CA 94305, USA}

\cortext[cor1]{Corresponding author}

\begin{abstract}
Experiments showing a seasonal variation of the nuclear decay rates of a number of different nuclei, and decay anomalies apparently related to  solar flares and solar rotation, have suggested that the Sun may somehow be influencing nuclear decay processes.  Recently, Cooper searched for such an effect in $^{238}$Pu nuclei contained in the radioisotope thermoelectric generators (RTGs) on board the Cassini spacecraft.  In this paper we modify and extend Cooper's analysis to obtain constraints on anomalous decays of $^{238}$Pu over a wider range of models, but these limits cannot be applied to other nuclei if the anomaly is composition-dependent.  We also show that it may require very high sensitivity for terrestrial experiments to discriminate among some models if such a decay anomaly exists, motivating the consideration of  future spacecraft experiments which would require less precision. 

\end{abstract}


\end{frontmatter}


\section{Introduction}
\label{Intro}

In a recent paper \cite{Messenger}, early data from a sample of $^{137}$Cs on board the {\sc Messenger} spacecraft enroute to Mercury were analyzed to set limits on a possible solar influence on nuclear decay rates.  This work was motivated by the  suggestion put forward in a recent series of papers which cite evidence for a drop in the count rate of $^{54}$Mn during a solar flare \cite{Flare}, for a correlation between decay rates of various isotopes and Earth-Sun distance \cite{Correlations,Space Science Review,BNL,PTB,Falkenberg,Parkhomov,Baurov,Ellis,Shnoll,Lobashev}, and for periodicities in decay-rate data associated with solar rotation \cite{Solar Physics,CPT 10}.  Although the suggestion of a solar influence on nuclear decay rates has been challenged by the apparent absence of decay anomalies in some isotopes that have been studied \cite{Norman,Cooper,Hardy}, and by a recent reactor experiment \cite{deMeijer}, there is no {\em a priori} reason to assume that all isotopes should be equally sensitive to a putative solar influence, or that  the antineutrinos produced in reactors would be the dominant agents through which a solar influence would be exerted.  As we have noted elsewhere \cite{Space Science Review,Semkow Response,Lindstrom II}, the very same properties of decaying nuclides that are responsible for the broad range of observed half-lives (e.g., nuclear and atomic wavefunctions, $Q$-values, selection rules) would likely render nuclides sensitive in different degrees to a putative solar influence.

The suggestion that the Sun is  responsible for  variations in decay rates  can be tested directly by studying the decay rates of appropriate nuclides located on spacecraft traveling through the solar system.  While such a specifically-designed mission has yet to be carried out,  a number of spacecraft have been launched to date carrying radioactive nuclides that can be used to constrain decay anomalies. As mentioned above,  Ref.~\cite{Messenger} develops a general formalism for constraining decay anomalies for nuclides placed on board spacecraft, such as the  $^{137}$Cs source on board {\sc Messenger}.  Additionally, as noted by Cooper \cite{Cooper}, the radioactive nuclides (e.g.,  $^{238}$Pu) used to generate electrical power  on  spacecraft like Cassini via radioisotope thermoelectric generators (RTGs) can also be used to set limits on decay anomalies.   The goal of this paper is to apply the general formalism developed in Ref.~\cite{Messenger} to spacecraft-borne RTGs.

The outline of our paper is as follows:  In Sec.~\ref{General Formalism section} we present the general formalism that develops the phenomenology characterizing a wide range of models of solar-induced decay anomalies, and show why experiments on board spacecraft may be crucial in studying them.  In Sec.~\ref{RTG section} we describe how this phenomenology can  be applied to obtain limits from spacecraft-borne RTGs.  In Sec.~\ref{Cassini section} we apply our formalism to the Cassini mission, modifying and extending Cooper's work \cite{Cooper} to a wider class of models.  We conclude in Sec.~\ref{Discussion section} with a summary and discussion of our results.

\section{General Formalism}
\label{General Formalism section}

We begin by briefly reviewing the formalism developed in Refs.~\cite{Messenger,Space Science Review} to describe anomalous radioactive decays.  If $N(t)$ is the number of unstable nuclei in a sample located at position $\vec{r}$, then we will assume that the total activity of the sample is 
\begin{equation}
-\frac{dN(t)}{dt} = \lambda(t)N(t) = \left[\lambda_{0} + \lambda_{1}(\vec{r},t)\right]N(t),
\label{N dot 1}
\end{equation}
where $\lambda_{0}$ represents the intrinsic contribution to the  decay rate of the unstable nuclei, along with a possible time-independent background arising from new interactions.  Here $\lambda_{1}(\vec{r},t)\ll \lambda_{0}$ characterizes the anomalous position- and time-dependent contribution to the decay rate,  assumed, in this case, to arise from the Sun.  

As noted above, there is evidence using terrestrial radioactive samples  suggesting that decay rates are correlated with sample-Sun separation.  To study this effect,  we will assume a specific phenomenological form for $\lambda_{1}(\vec{r},t)$ given by~\cite{Messenger}
\begin{equation}
\lambda_{1}(\vec{r},t) = \lambda_{0}\xi^{(n)}\left[\frac{R}{r(t)}\right]^{n},
\label{lambda 1}
\end{equation}
where $n = 1, 2, 3, \ldots$,   $r = |\vec{r}\,|$ is now the sample-Sun separation distance, $R = 1$~AU = $1.495979\times 10^{8}$~km, and $\xi^{(n)}$ is a composition-dependent dimensionless parameter characterizing the ``strength'' of the decay anomaly for a specific nucleus.   The form of Eq.~(\ref{lambda 1}) is designed to encompass a broad range of theories.  If the decay anomalies are caused by a flux of neutrinos  from the Sun or by an inverse-square law field, then one expects 
$n = 2.$  On the other hand, radioactive decay processes might be affected by one of the  many proposed new long-range inverse-power-law interactions.  In this case, the potential energy  between point particles of mass $m_{1}$ and $m_{2}$ separated by distance $r$ may be written in the general  form \cite{IPL}
\begin{equation}
V_{n}(r) = -\alpha_{n}\left(\frac{Gm_{1}m_{2}}{r}\right)\left(\frac{r_{0}}{r}\right)^{n -1},
\label{Vn}
\end{equation}
where $\alpha_{n}$ is a dimensionless constant, $G$ is the Newtonian gravitation constant, and $r_{0} = 10^{-15}$~m is a length scale chosen by convention.  For example, $n = 2$ can arise from the simultaneous exchange of 2 massless scalar particles \cite{Sucher}, $n = 3$ from 2-massless-pseudoscalar exchange \cite{Most Sokolov,Ferrer}, and $n = 5$ from 2-neutrino \cite{Feinberg,Fischbach} and 2-axion exchanges \cite{Ferrer}.

Because the Earth-Sun variation of $r$ is small, it is difficult to obtain the dependence on the power $n$ from a purely terrestrial experiment, which is why the use of  spacecraft with widely varying values of $r$ is important.   To show this, we
substitute Eq.~(\ref{lambda 1}) into Eq.~(\ref{N dot 1}), giving
\begin{equation}
-\frac{dN(t)}{dt} = \lambda(t)N(t) =\lambda_{0} \left\{1 +\xi^{(n)}\left[\frac{R}{r(t)}\right]^{n}\right\}N(t).
\label{N dot 2}
\end{equation}
This form is not very useful since it depends on the instantaneous value of $N(t)$.  Following Ref.~\cite{Messenger}, we can eliminate this dependence by first integrating Eq.~(\ref{N dot 2}), obtaining
\begin{equation}
N(t) = N_{0}\exp\left\{-\lambda_{0}\left[t + \xi^{(n)} {\cal I}^{(n)}(t)\right]\right\},
\label{N(t)}
\end{equation}
where $N_{0} = N(0)$ and 
\begin{equation}
{\cal I}^{(n)}(t) = \int^{t}_{0}dt'\left[\frac{R}{r(t')}\right]^{n}.
\label{I n}
\end{equation}
Then differentiating Eq.~(\ref{N(t)}) gives
\begin{equation}
-\dot{N}(t)= \lambda_{0}N_{0}\exp(-\lambda_{0}t)\left\{1 + \xi^{(n)} \left[\frac{R^{n}}{r^{n}(t)} - \lambda_{0}{\cal I}^{(n)}(t)\right]\right\},
\label{N dot 3}
\end{equation}
where $\dot{N}(t) \equiv dN(t)/dt$ and assuming  $\xi^{(n)}\lambda t\ll 1$,.  For our application, we will set $t = 0$ to be the launch time of the spacecraft, in which case $r(0) = R.$  Then, using ${\cal I}^{(n)}(0) = 0$, Eq.~(\ref{N dot 3}) yields
\begin{equation}
-\dot{N}(0)= \lambda_{0}N_{0}\left[1 + \xi^{(n)}\right].
\label{N dot t = 0}
\end{equation}

For practical purposes, it is useful to express Eq.~(\ref{N dot 3}) in terms of directly measurable quantities.  If we
define $\lambda \equiv \lambda_{0}[1 + \xi^{(n)}]$ to be the decay rate  observed on Earth when $r = R$, then $\lambda_{0} = \lambda[1 - \xi^{(n)}] + {\cal O}[\xi^{(n)}]^{2}.$  Substituting this into Eq.~(\ref{N dot 3}) and retaining only  terms of leading order in $\xi^{(n)}$,  we find after using Eq.~(\ref{N dot t = 0}) that \cite{Messenger}
\begin{equation}
\dot{N}(t) \simeq \dot{N}(0)e^{-\lambda t}\left[1 + \xi^{(n)} B^{(n)}(t)\right],
\label{N dot}
\end{equation}
where
\begin{equation}
B^{(n)}(t) \equiv \left\{\left[\frac{R}{r(t)}\right]^{n} - 1\right\} - \lambda\left[ {\cal I}^{(n)}(t) - t \right].
\label{B}
\end{equation}
For later purposes, we note from Eq.~(\ref{B}) that $B^{(n)}(0) = 0.$  The  terms in braces in Eq.~(\ref{B}) represent the anomalous contributions to $\dot{N}(t)$ arising from the circumstance that at some $t$ the sample is at $r(t)$, rather than at $R$.  The second term  in square brackets represents an additional {\em cumulative contribution} to $\dot{N}(t)$ from the variation of $r(t)$ from $t = 0$ to $t$, and this is generally nonzero even when $r(t) = R$ (see below).

We will now demonstrate why  spacecraft missions may be needed to distinguish among the inverse-power-law models given by Eq.~(\ref{lambda 1}).  Since the eccentricity of the Earth's orbit is small ($\varepsilon_{\oplus} \simeq 0.0167$),  the Earth-Sun separation can be written as
\begin{equation}
r_{\oplus}(t) = \overline{r}_{\oplus} + \varepsilon_{\oplus} \overline{r}_{\oplus}\sin\left(\frac{2\pi t}{T}\right) + {\cal O}(\varepsilon_{\oplus}^{2}),
\label{r earth}
\end{equation}
where $ \overline{r}_{\oplus} \simeq R$ is the mean Earth-Sun separation, $T = 1$ year is the orbital period, and $t =0$ when the Earth is at $r = R$.   Substituting Eq.~(\ref{r earth}) into Eq.~(\ref{I n}) and integrating, while keeping only terms of ${\cal O}(\varepsilon_{\oplus})$, we find
\begin{equation}
{\cal I}^{(n)}_{\oplus}(t) \simeq t - \frac{n\varepsilon_{\oplus}T}{2\pi}\left[1 - \cos\left(\frac{2\pi t}{T}\right)\right],
\end{equation}
so $B^{(n)}(t)$ given by Eq.~(\ref{B}) becomes, after neglecting terms of ${\cal O}(\varepsilon_{\oplus}^{2}),$
\begin{equation}
B^{(n)}_{\oplus}(t) \simeq -n\varepsilon_{\oplus}\left\{\sin\left(\frac{2\pi t}{T}\right) - \frac{\lambda T}{2\pi}\left[1 -\cos\left(\frac{2\pi t}{T}\right)\right]\right\}.
\label{B earth}
\end{equation}
Combining Eqs.~(\ref{N dot}) and (\ref{B earth}), we find that the activity of a sample on Earth is given by
\begin{eqnarray}
\dot{N}_{\oplus}(t) &\simeq& \dot{N}_{\oplus}(0)e^{-\lambda t}
	\left(1 - n\xi^{(n)}\varepsilon_{\oplus}
		\left\{\sin\left(\frac{2\pi t}{T}\right) 
		\right.\right.\nonumber\\
		& & \mbox{}\left.\left.- \frac{\lambda T}{2\pi}\left[1 - \cos\left(\frac{2\pi t}{T}\right)\right]\right\}\right).
\label{N dot earth}
\end{eqnarray}
Since $n$ and $\xi^{(n)}$  appear together only in the combination $n\xi^{(n)}$  in Eq.~(\ref{N dot earth}), it is not possible   to distinguish between theories with different values of  $n$ using terrestrial decay experiments unless one has sufficient sensitivity to detect effects of ${\cal O}(\varepsilon_{\oplus}^{2}) \sim 3 \times 10^{-4}$.  On the other hand, by placing  radioactive samples on board spacecraft,  a decay anomaly can be probed over a wide range of $r(t)$, allowing one to  discriminate more easily between the various powers of $n$ given in Eq.~(\ref{lambda 1}).

\section{Application to RTGs}
\label{RTG section}

The thermoelectric generators on board spacecraft use the Seebeck effect to convert heat from a hot reservoir into electrical power, while waste heat is exhausted into a cold reservoir.  In a simple model \cite{Angelo},
the total electrical power output can be written as 
\begin{equation}
P_{\rm el}(t) = I\Delta V_{\rm term},
\label{P el}
\end{equation}
where $I$ is the current, and the terminal voltage across the generator is given by
\begin{equation}
\Delta V_{\rm term} = \alpha_{n,p}(T_{H} - T_{C}) - IR_{\rm int}.
\end{equation}
Here $\alpha_{n,p}$ is the differential Seebeck coefficient for the $n$- and $p$-doped semiconductor generator legs, $T_{H}$ and $T_{C}$ are the temperatures of the hot and cold heat reservoirs, and $R_{\rm int}$ is the internal resistance of the generator.    The efficiency of a thermoelectric generator is usually expressed as
\begin{equation}
\eta_{TG} = \frac{P_{\rm el}}{\dot{Q}_{H}},
\end{equation}
where $\dot{Q}_{H}$ is the thermal power input from the hot reservoir.

An RTG is a thermoelectric generator that uses a radioactive material as its hot temperature reservoir.
RTGs have been placed on board spacecraft since the early 1960s \cite{Angelo,Hyder}, and  are typically used on probes traveling outward in the solar system where solar panels do not provide sufficient power.  These missions include Pioneer 10 and 11, Voyager 1 and 2, Galileo, Ulysses, Cassini, and New Horizons.  In addition, RTGs were used to power instruments on the two Viking landers on Mars, and instruments on the Moon placed by  Apollo astronauts.

The thermal power $P_{\rm th}(t)$ generated by the radioactive heat source of an RTG is directly proportional to $\dot{N}(t)$,  the activity of the radioactive material used.  Therefore,  we can use Eq.~(\ref{N dot}) to relate the thermal power generated at time $t$ to the thermal power produced at launch $t = 0$:
\begin{equation}
P_{\rm th}(t) = P_{\rm th}(0)e^{-\lambda t}\left[1 + \xi^{(n)} B^{(n)}(t)\right].
\label{thermal power}
\end{equation}
Unfortunately for our purposes, this thermal power is not directly observed.  Instead, the electrical power $P_{\rm el}(t)$ given by Eq.~(\ref{P el}) is measured.   To relate $P_{\rm el}(t)$ to $P_{\rm th}(t)$, we  follow Cooper \cite{Cooper} and introduce the dimensionless efficiency function $\epsilon(t)$ defined by
\begin{equation}
P_{\rm el}(t) \equiv \epsilon(t)P_{\rm th}(t),
\end{equation}
where 
\begin{equation}
\epsilon(t) = \eta_{TG}(t)\eta_{\rm rad}(t),
\end{equation}
and 
\begin{equation}
\eta_{\rm rad} \equiv \frac{\dot{Q}_{H}}{P_{\rm th}}
\end{equation}
is the fraction of the total radioactive thermal power that flows into the generator.
Eq.~(\ref{thermal power}) can then be rewritten as
\begin{subequations}
\label{electrical power}
\begin{align}
P_{\rm el}(t) &=   P_{\rm el}(0)\,\epsilon(t)\,e^{-\lambda t}\left[1 + \xi^{(n)} B^{(n)}(t)\right],  
\label{P el lambda} \\
		&=   P_{\rm el}(0)\,\epsilon(t)\,2^{-t/t_{1/2}}\left[1 + \xi^{(n)} B^{(n)}(t)\right],
\label{P el halflife} 
\end{align}
\end{subequations}
where $t_{1/2} = (\ln 2)/\lambda$ is the half-life of the nuclide.  We note that $\epsilon(t)$ is time dependent since  RTG efficiency generally decreases with time. 

For our problem, $\epsilon(t)$ must be determined in a manner that is not influenced by the presence of a decay anomaly, and one option is to use a computer model.  However, the model commonly used to describe RTG performance (DEGRA \cite{DEGRA}) assumes the
usual exponential decay of the radionuclide, preventing its use in
testing the exponential decay hypothesis \cite{Cooper}.   The alternative  suggested by Cooper \cite{Cooper} is to use an empirical approach that capitalizes on the fact that there may be points along a spacecraft's trajectory where anomalous decay effects make no contribution, which from Eq.~(\ref{electrical power}), occurs  whenever $B^{(n)}(t) = 0$.  In his analysis of Cassini power data, Cooper utilized five points at times where $r(t) = R$, but from Eq.~(\ref{B}) we see that
\begin{equation}
B^{(n)}(r = R) = -\lambda\left[{\cal I}^{(n)}(t) - t\right],
\end{equation}
which does not generally vanish for $t > 0$, though in the specific case of Cassini  studied by Cooper, this is a reasonable
approximation.    If we define the decay-normalized electrical power as
\begin{equation}
{\cal P}(t) \equiv 2^{+t/t_{1/2}}\frac{P_{\rm el}(t)}{P_{\rm el}(0)} = e^{+\lambda t}\frac{P_{\rm el}(t)}{P_{\rm el}(0)} ,
\label{cal P}
\end{equation}
and use only power data where $B^{(n)}(t)  = 0$, then we can solve  Eq.~(\ref{electrical power}) for $\epsilon(t)$ in terms of known quantities,
\begin{equation}
\epsilon(t) = {\cal P}(t; B^{(n)} = 0),
\label{finding e}
\end{equation}
and fit the results to an empirical function as we will demonstrate in the next section.

This analysis indicates that decay anomalies can, in principle, be detected using the electrical power output as a proxy for the (time-dependent) energy release from nuclear decays in spacecraft RTGs.  All that is required is: (1) a spacecraft trajectory with a significant variation of $r(t)$ (preferably including  $r(t) \ll R$), 
(2) accurate  measurements of the spacecraft's position and electrical power production, and (3) a good model of the RTG efficiency function $\epsilon(t)$ that does not assume the radioactive decay law.  If a computer model for the efficiency is unavailable, an empirical
approach can be used, provided there are a sufficient number of points
where $B(t) = 0$.    The original Solar Probe Plus  mission,  which would have used a spacecraft  powered by RTGs while using a Jupiter gravity assist maneuver to place it into solar polar orbit, would have been ideal for our purposes \cite{Solar Probe}.  However, this planned mission has been changed to avoid the use of RTGs.  Of the  interplanetary missions  that have used RTGs, only Cassini and Galileo actually crossed  the 1-AU orbit radius after launch, allowing  the empirical approach.    Since the Cassini mission crossed the most times, it provides the most complete data to model the RTG efficiency empirically and so will be used in this paper.

We note in passing that accurate modeling of RTGs has also been important in understanding the Pioneer anomaly, the small anomalous acceleration of both Pioneer 10 and 11 spacecraft  \cite{Turyshev LR}.   In order to explain the observed temporal decay of this acceleration,  the changes in thermal recoil forces on these spacecraft due to the degradation of the RTGs need to be accurately modeled \cite{Turyshev PRL}.

\section{Application to Cassini RTG Data}
\label{Cassini section}

Launched in 1997, the Cassini spacecraft made several gravity-assisted flybys of Venus,  Earth, and Jupiter  before  entering the Saturn system in 2004.  In principle, this nearly 7-year-journey provides both a sufficiently long duration and a substantial variation in $r(t)$ to render Cassini  ideal for a test of solar-influenced radioactive decays.  However, following  Cooper \cite{Cooper}, we will only use the first two years of data, when 0.6732~AU $\leq r \leq$ 1.6215~AU, because only within this range can we determine with some confidence the efficiency function $\epsilon(t)$.  For $t > 2$~years, there are no longer any times where $B^{(n)}(t) = 0$ which can be used to perform the empirical fit to determine $\epsilon(t)$.  

The Cassini spacecraft uses three  General Purpose Heat Source Radioisotope Thermal Generators (GPHS-RTGs), each producing nearly 300~W of electrical power from 572 thermoelectric ``unicouples.''   A GPHS generates approximately 4410 W of thermal power at the beginning of life from PuO$_{2}$ fuel pellets enriched to about 80\% $^{238}$Pu, corresponding to about 8.1~kg of $^{238}$Pu per generator \cite{Daring}.    The pellets were formed from PuO$_{2}$ powder obtained from Russia, where the plutonium was created by irradiating $^{237}$Np in a high-flux reactor to form $^{238}$Np  which has a half-life of 2.4 days \cite{O'Brien}.  A $^{238}$Np nucleus  decays to $^{238}$Pu via a $\beta^{-}$-decay.  For the rest of this analysis, we will assume that all of the thermal heat production of a GPHS  results from $^{238}$Pu, which has a half-life $t_{1/2} = 87.7$~years.  (The other most likely nuclide present would be $^{239}$Pu, whose half-life is $2.41 \times 10^{10}$~yr, and hence decays too slowly to contribute significantly to the energy production in an RTG.)

\begin{figure}[tbp]
\begin{center}
\includegraphics[width=3in]{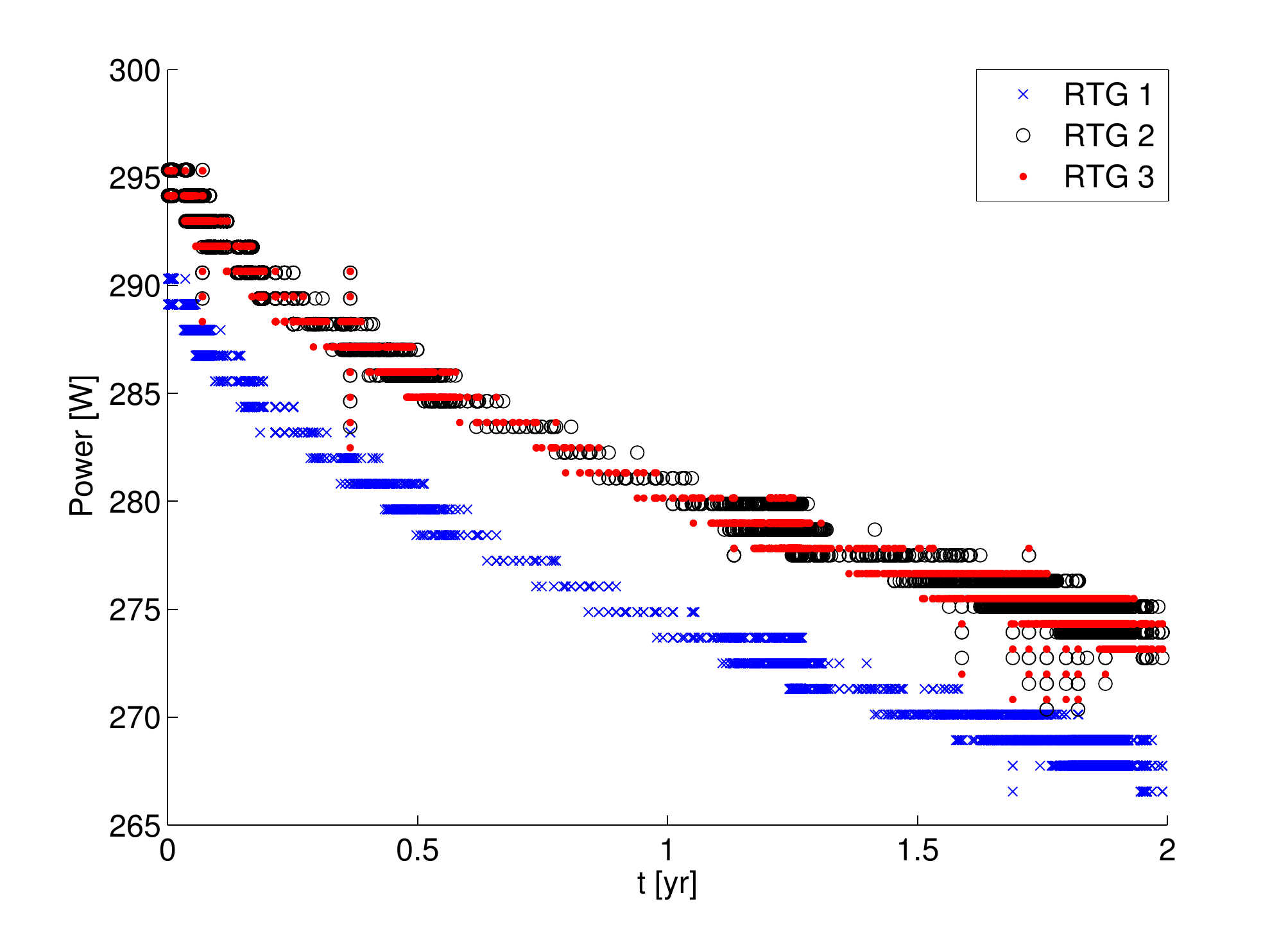}
\caption{Electrical power output from each of the three Cassini RTGs for the first two years of the mission.  The digitization arises from the current measurements.}
\label{Power figure}
\end{center}
\end{figure}

Our analysis begins with Cassini's trajectory and RTG current $I(t)$ data  provided by the Jet Propulsion Laboratory.  Since the current is monitored separately for each RTG, each can provide a separate determination of the material-dependent parameter $\xi_{\rm Pu}^{(n)}$ for $^{238}$Pu, and together the RTGs can be used to assess the validity of the empirical efficiency function $\epsilon_{\rm emp}(t)$.   A total of 157,465 current measurements (in increments of 0.03959~A) were taken over the 2-year period at irregular intervals, and a  small number of obviously spurious points were removed before our analysis began.   The Cassini power system  is regulated with a variable shunt radiator to maintain a constant terminal voltage $\Delta V_{\rm term} = 30.0\pm 0.2$~V  across each RTG  to maximize power production \cite{Ging}, and hence the  RTG electrical power was obtained from the current measurements using Eq.~(\ref{P el}).   Since the variable shunt regulator is located ``downstream'' from the RTG, its operation does not affect our ability to determine the RTG's power output as a function of time.  The results are shown in Fig.~\ref{Power figure} with digitization of the current measurements clearly evident.
The power decreases more rapidly than expected from the radioactivity exponential decay law for $^{238}$Pu due to unicouple degradation  among other factors.

\begin{figure}[tbp]
\begin{center}
\includegraphics[width=3in]{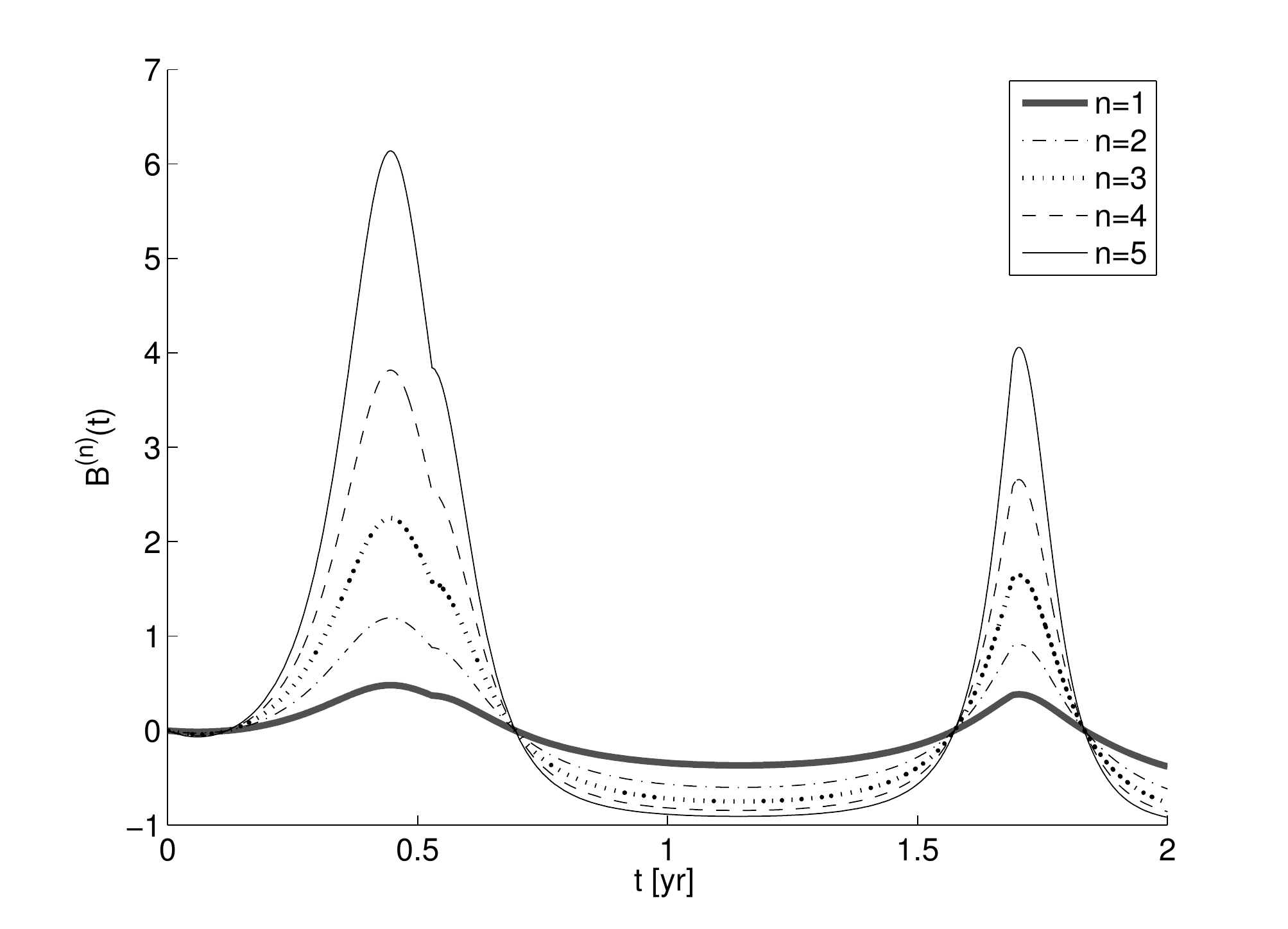}
\caption{$B^{(n)}(t)$ of Cassini obtained using Eq.~(\ref{B}) for the first two years of the mission.  The five points where $B^{(n)}(t) = 0$, which occur when $r \simeq R$,  were used to obtain the empirical efficiency function as shown in Fig.~\ref{efficiency figure}.}
\label{B figure}
\end{center}
\end{figure}

Using the Cassini trajectory data, $B^{(n)}(t)$ for $n = 1$--5 was computed using Eq.~(\ref{B}) for $0 \leq t \leq 2$~years, and is plotted in Fig.~\ref{B figure}.
We  note that $B^{(n)}(t)$ increases with $n$, so 
the constraints on $\xi^{(n)}_{\rm Pu}$ become more stringent as $n$ increases. From Eq.~(\ref{B}), we see that the largest values of $B^{(n)}(t)$, which  lead to the largest decay anomalies, occur when $r(t)$ is smallest.  For Cassini, this occurs at its closest approach to the Sun, $r = 0.6732$~AU at $t \simeq 0.45$~yr, with smaller peaks at $t \simeq 1.7$~yr.   We also note from Fig.~\ref{B figure} that there are five points where $B^{(n)}(t) = 0,$ which will be needed to obtain the empirical efficiency function.

\begin{figure}[tbp]
\begin{center}
\includegraphics[width=3in]{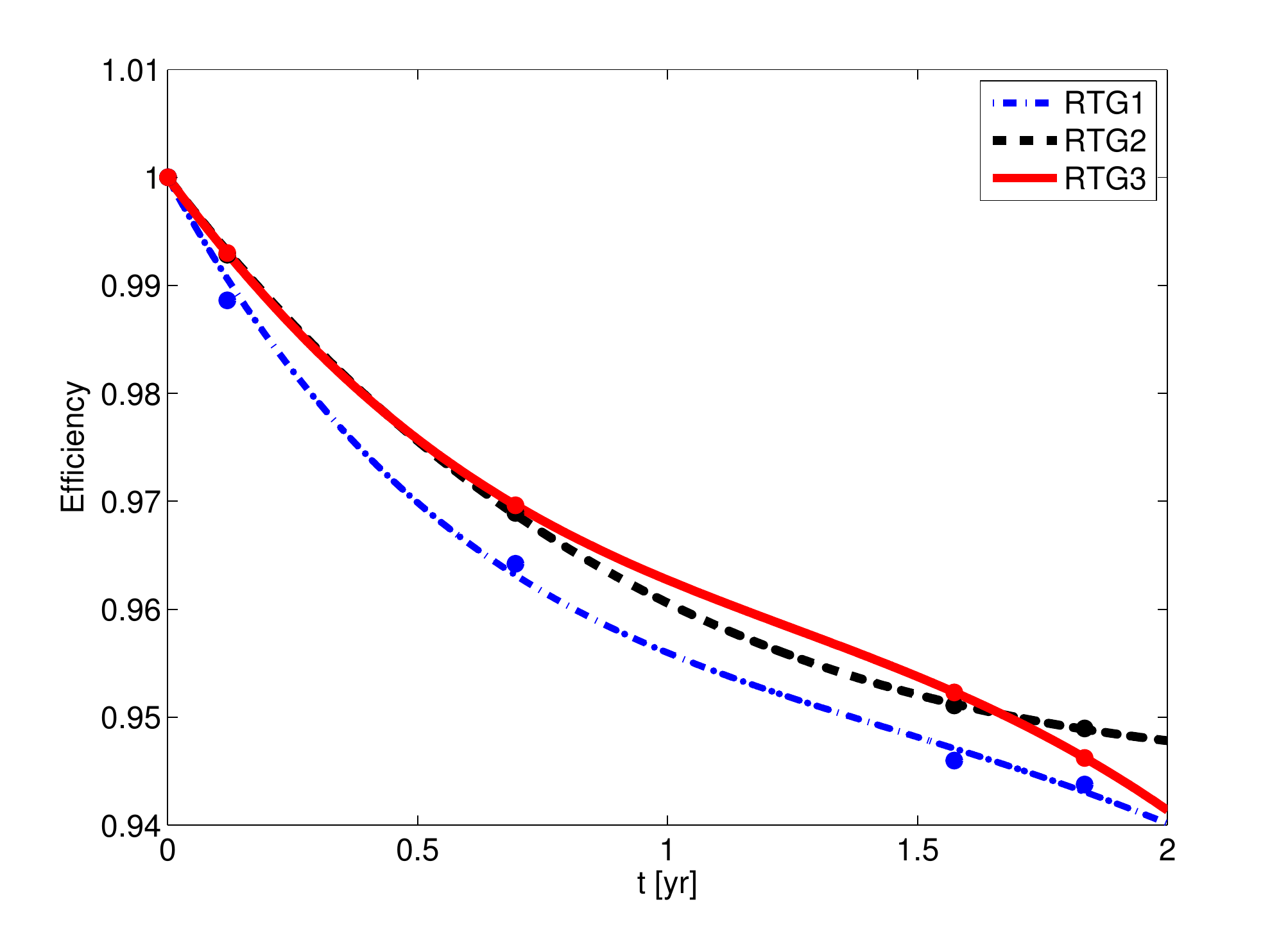}
\caption{Fits of ${\cal P}(t)$ where $B^{(n)}(t) = 0$ to $ \epsilon_{\rm emp}(t)$ given by Eq.~(\ref{e emp}) for the three Cassini RTGs for $n = 2$.  A small variation in efficiency among the RTGs is observed. }
\label{efficiency figure}
\end{center}
\end{figure}

Using the data from each of the three RTGs for the first two years, we calculated the decay-normalized power ${\cal P}(t)$ using Eq.~(\ref{cal P}).  For the 5 data points where $B^{(n)}(t) = 0$, we then determined $\epsilon(t)$ following Eq.~(\ref{finding e}) by plotting  for each RTG  $\epsilon(t) = {\cal P}(t; B = 0)$ versus $t$, and then fitting the results to an empirical function similar to that suggested by Cooper \cite{Cooper},
 \begin{equation}
 \epsilon_{\rm emp}(t) = 2^{-t/T_{\rm eff}}\left(1 + at + bt^{2} + ct^{3}\right),
 \label{e emp}
 \end{equation}
 where $T_{\rm eff}$, $a$, $b$, and $c$ are constants.  [Note that our procedure differs from Cooper who only used $\epsilon_{\rm emp}(t) = 2^{-t/T_{\rm eff}}$ for the points where $r(t) =R$.  The third-order polynomial portion of Eq.~(\ref{e emp}) was  introduced later and fit using all of the trajectory data $0 \leq t \leq 2$ yr.]  The result for $n = 2$ is shown in Fig.~\ref{efficiency figure}; the results for the other values of $n$ are virtually identical.
 We see that  there is some relative variation among the three RTGs, and we use this variation to estimate the relative uncertainty in the empirical efficiency function to be $\delta\epsilon_{\rm emp}/\epsilon_{\rm emp} \lesssim 0.5\%$.
 
\begin{figure}[tbp]
\begin{center}
\includegraphics[width=3in]{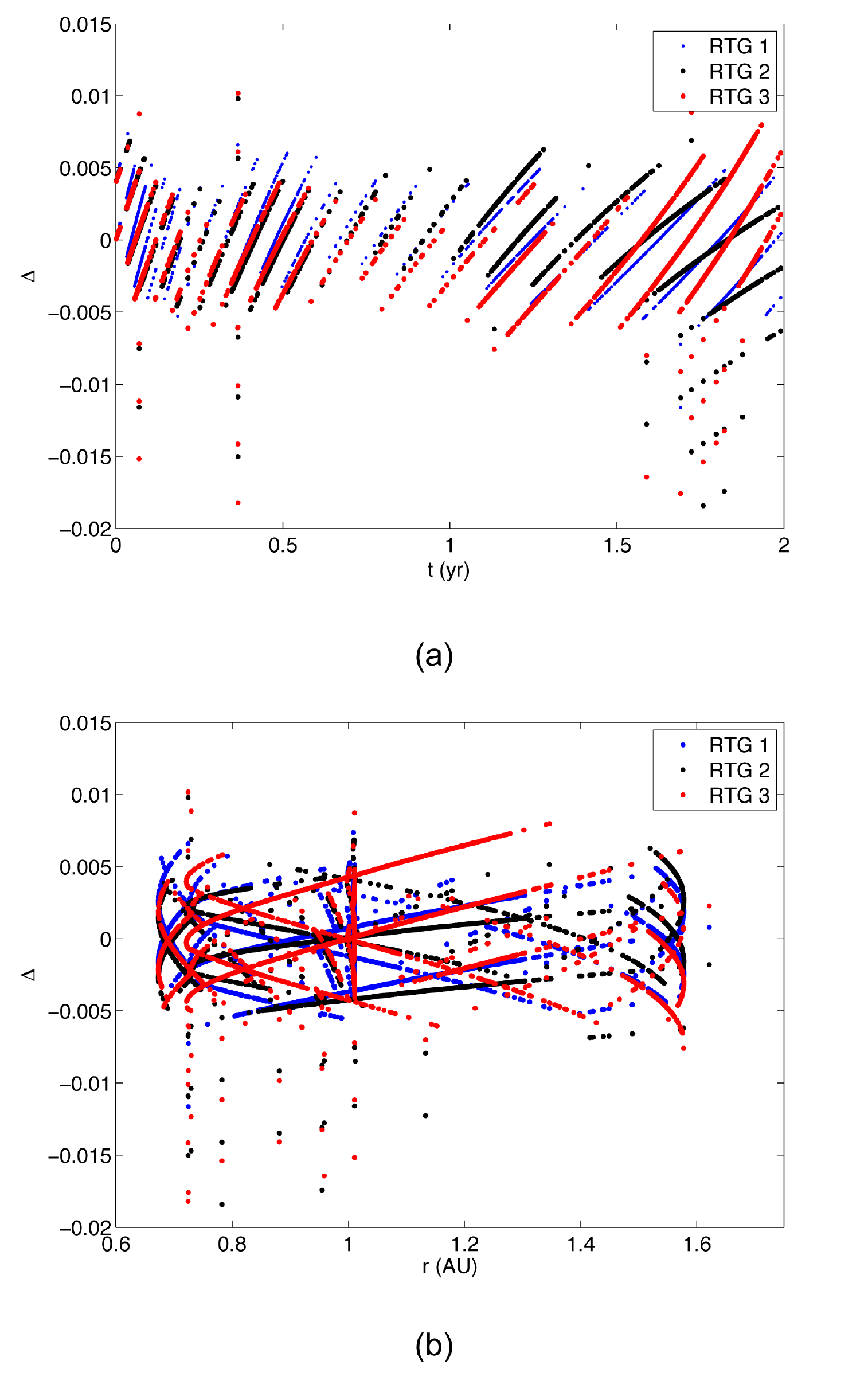}
\caption{$\Delta$ given by Eq.~(\ref{delta}) plotted versus $t$ (a) and $r$ (b).  The striped patterns are due to the digitization of the power seen in Fig.~\ref{Power figure}.  No other significant systematic pattern is observed.}
\label{delta figure}
\end{center}
\end{figure}

 We can assess the validity of our empirical approach for determining $\epsilon(t)$ by examining the quantity
 \begin{equation}
 \Delta(t) \equiv \left[\frac{{\cal P}(t)}{\epsilon_{\rm emp}(t)} -1\right],
 \label{delta}
 \end{equation}
 which would vanish if $\epsilon_{\rm emp}(t)$ gave a perfect characterization of ${\cal P}(t)$.  In Fig.~\ref{delta figure}, $\Delta$ is plotted versus $t$ and $r$.
Using the combined data from all three RTGs, we  find  
\begin{equation}
|\Delta(t)| \lesssim 0.005,
\label{Delta value}
\end{equation}
 with no significant systematic  dependence of $\Delta$ on $t$ or $r$ other than the effects due to the digitization of the power seen in Fig.~\ref{Power figure}.   Thus, $\epsilon_{\rm emp}(t)$ gives a good characterization of the decay-normalized power of the RTGs to the level of 0.5\%.  Systematic effects due to solar heating of the RTGs, which would depend on $1/r^{2}$, are not evident.   The semiconductor legs of the unicouples generating the power via the temperature difference $T_{H} - T_{C}$  are only 20.3~mm long \cite{Daring}.  Only a differential heating over this distance scale would produce a change in the RTG electrical power production, which is not seen.

Now that the efficiency function $\epsilon(t) = \epsilon_{\rm emp}(t)$ has been determined, all quantities in Eq.~(\ref{electrical power}) are known except for $\xi^{(n)}_{\rm Pu}$.   Inserting Eq.~(\ref{cal P}) into Eq.~(\ref{electrical power}) and setting $\epsilon(t) \simeq \epsilon_{\rm emp}(t)$, we can solve for $\xi^{(n)}$ to obtain
\begin{equation}
 \xi^{(n)}_{\rm Pu}  = \frac{\Delta(t)}{B^{(n)}(t)},
\label{simple xi}
\end{equation}
where we have used Eq.~(\ref{delta}).  Inserting the numerical limit on $\Delta(t)$ given by Eq.~(\ref{Delta value}) into Eq.~(\ref{simple xi}), we find
\begin{equation}
\left| \xi^{(n)}_{\rm Pu}\right|  \lesssim \frac{0.005}{|B^{(n)}(t)|}.
\label{xi limit}
\end{equation}
The most stringent limits on $\xi^{(n)}_{\rm Pu}$, which are obtained  when $|B^{(n)}(t)| = B^{(n)}_{\rm max}$,  the maximum observed value of $B^{(n)}(t)$, are presented in Table~\ref{xi table}.
\begin{table}[tdp]
\caption{The constraints on  $\xi^{(n)}_{\rm Pu}$ for $^{238}$Pu obtained from  Eq.~(\ref{xi limit}) when $|B^{(n)}(t)| = B^{(n)}_{\rm max}$. }
\begin{center}
\begin{tabular}{ccc} \hline
$n$ & $B^{(n)}_{\rm max}$ & $|\xi_{\rm Pu}^{(n)}|$ \\ \hline
1 & 0.48 & $< 1.0 \times 10^{-2} $ \\
2 & 1.19 & $<  4.2 \times 10^{-3}$\\
3 & 2.25 &$< 2.2 \times 10^{-3}$ \\
4 & 3.82 & $< 1.3 \times 10^{-3}$\\
5 & 6.14 &$< 8.1 \times 10^{-4}$ \\ \hline
\end{tabular}
\end{center}
\label{xi table}
\end{table}%
We see that the tightest constraints are obtained for the largest values of $n$, which is consistent with our earlier discussion, while a comparitively  poor constraint is obtained for $n = 1$ since $B^{(1)}_{\rm max}$ is relatively small.

\section{Discussion}

From our analysis of the Cassini RTG power data, we find no evidence of a solar-influenced decay anomaly for $^{238}$Pu having the inverse-power-law form given by Eq.~(\ref{lambda 1}) for $n = 1$--5.    Unfortunately, it is difficult to compare our constraints on  $\xi_{\rm Pu}^{(n)}$ given in Table~\ref{xi table} with the earlier analysis by by Cooper \cite{Cooper}, who also found no anomaly.  Schematically, Cooper fit the total normalized RTG power production to an empirical efficiency function, and then fit the resulting residuals  to two different possible functions characterizing the decay anomaly:
\begin{equation}
\mbox{residuals} = \left\{\begin{array}{c}
				\displaystyle \alpha\left[\frac{R}{r}\right]^{2}, \\
				\\
				\displaystyle \beta\left[\frac{R}{r}\right],
				\end{array}
				\right.
\label{Cooper}
\end{equation}
where $R =$ 1 AU.   This procedure yielded the 90\% confidence level limits
\begin{subequations}
\label{Cooper's limits}
\begin{align}
|\alpha| &<  0.84 \times 10^{-4},\\ 
|\beta| &<  0.99 \times 10^{-4}.
\end{align}
\end{subequations}
 However, despite their superficial similarity, Cooper's formulas given by Eq.~(\ref{Cooper}) are not directly related to our more fundamental formula for $\lambda_{1}$ given by Eq.~(\ref{lambda 1}).  The anomaly in the power production of the RTGs in our approach is not a simple power law since one needs to take into account the fact that the sample's activity at time $t$ actually depends on the sample's {\em entire history} because of  its dependence on the function ${\cal I}^{(n)}(t)$ given by Eq.~(\ref{I n}).  Since the two approaches of characterizing the power anomaly are so different,  it is difficult to relate Cooper's limits on $\alpha$ and $\beta$ given by Eq.~(\ref{Cooper's limits}) to our constraints on $\xi_{\rm Pu}^{(2)}$ and $\xi_{\rm Pu}^{(1)}$, respectively, given in Table~\ref{xi table}.

As noted in the Introduction,  one expects an anomalous decay mechanism to depend on the nuclei and decay process so one cannot, without additional assumptions,  use the limits on $\xi^{(n)}$ for $^{238}$Pu to constrain anomalous decays of  other nuclides or to refute the observations of previous experiments. The Cassini RTGs are powered  exclusively by alpha-decays due to the extremely long half-lives of the uranium daughters, particularly $^{234}$U ($t_{1/2} = 246,000$~yr) which is the first daughter of $^{238}$Pu.  The absence of a decay anomaly for  $^{238}$Pu contrasts with the other isotopes in which an anomaly has  potentially been observed \cite{Correlations,Space Science Review,BNL,PTB,Falkenberg,Parkhomov,Baurov,Ellis,Shnoll,Lobashev,Solar Physics,CPT 10}, that are beta-decays or (like $^{226}$Ra) where one actually measures a significant beta-decay component of daughter products \cite{Buncher}.   For example, Parkhomov \cite{Parkhomov},  found time-dependent fluctuations in $^{90}$Sr, $^{90}$Y, and $^{60}$Co, all of which are beta decays, but not in $^{239}$Pu which, like $^{238}$Pu studied in this paper,  is an alpha decay.  Each radioactive nuclide needs to be examined separately for anomalous decay processes, and  $\xi^{(n)}$ determined for each.   If Eq.~(\ref{lambda 1}) is correct, all anomalous effects should be characterized by the same power $n$, and all experiments using the same nuclide should yield the same value of $\xi^{(n)}$.   Work is currently under way to apply  Eq.~(\ref{lambda 1}) to the results of terrestrial experiments, as described in Sec.~2.   The only previously reported result is  $\xi^{(2)}_{\rm Cs} = (2.8 \pm 8.1)\times 10^{-3}$ for the $^{137}$Cs sample aboard the Messenger spacecraft \cite{Messenger}.

We have also shown that if an inverse-power-law form given by Eq.~(\ref{lambda 1}) exists, terrestrial  decay experiments will require unusually high sensitivity to discriminate $\xi^{(n)}$ from $n$ since only the combination $n\xi^{(n)}$ appears to first order in the orbital eccentricity.  Thus, experiments  conducted on board spacecraft may be needed to distinguish among the various powers of $n$.   While we and Cooper \cite{Cooper} have shown that spacecraft-borne RTGs can be used, the isotopes used for power generation are very limited, and hence dedicated experiments using nuclides that have demonstrated a decay anomaly in terrestrial experiments should be used.

Since RTGs will continue to be a power source on board spacecraft, improved instrumentation on future missions could not only be used to refine RTG performance models, but also provide improved tests of the radioactivity exponential decay law.   The two major factors limiting constraints on decay anomalies using the Cassini RTGs are the uncertainties of the empirical efficiency model and the resolution of the electric current (and hence, power) measurements.  Additional thermal measurements would also be useful in allowing a more direct determination of the thermal power output of the RTG heat source.

\label{Discussion section}

\section*{Acknowledgements}

The work of E.F. is supported in part by USDOE contract no. DE-AC02-76ER071428.  Part of this research was carried out at the Jet Propulsion
Laboratory, California Institute of Technology, under a contract with the National Aeronautics and Space Administration.



\bibliographystyle{elsarticle-num}
\bibliography{<your-bib-database>}



\end{document}